\documentstyle[12pt]{article}
\evensidemargin \oddsidemargin
\catcode `\@=11
\def\numberbysection{\@addtoreset{equation}{section}
         \renewcommand{\theequation}{\thesection.\arabic{equation}}}
\def\be{\begin{equation}}
\def\ee{\end{equation}}
\newcommand{\ba}{\begin{eqnarray}}
\newcommand{\ea}{\end {eqnarray}}
\newcommand{\nn}{\nonumber}

\newcommand{\bulksp}[2]{\langle#1\,|\,#2\rangle_{\mbox{\scriptsize{bulk}}}}
\newcommand{\randsp}[2]{\langle#1\,|\,#2\rangle
_{\mbox{\scriptsize{boundary}}}}
\def\a{\alpha}
\def\b{\beta}

\def\d{\delta}

\def\e{\eta}

\def\L{\Lambda}

\def\p{\phi}

\def\t{\tau}

\def\O{\Omega}

\def\w{\omega}
\def\nil{\emptyset}

\def\inf{\infty}
\def\half{\frac{1}{2}}

\numberbysection
\begin{document}
%
\pagestyle{empty}
\vspace* {5mm}
\renewcommand{\thefootnote}{\fnsymbol{footnote}}
\begin{center}
\large
   {\bf FINITE-SIZE SCALING STUDIES}\\[4mm]
   {\bf OF REACTION-DIFFUSION SYSTEMS}\\[4mm]
   {\bf Part II: Open Boundary Conditions}\\[1cm]
\end{center}
\begin{center}
\normalsize
  Haye Hinrichsen$^{\dag}$, Klaus Krebs$^{\ddag}$,
  Markus Pfannm\"uller$^{\ddag}$ and Birgit Wehefritz$^{\ddag}$ \\[5mm]
  $^{\dag}$ {\it Freie Universit\"{a}t Berlin, Fachbereich Physik\\
    Arnimallee 14, D-14195 Berlin, Germany}\\[4mm]
  $^{\ddag}$ {\it Universit\"{a}t Bonn,
   Physikalisches Institut \\ Nu\ss allee 12,
   D-53115 Bonn, Germany}
\\[15mm]
{\bf Abstract}
\end{center}
\renewcommand{\thefootnote}{\arabic{footnote}}
\addtocounter{footnote}{-1}
%
%
\small

We consider the coagulation-decoagulation model on an one-dimensional lattice
of length $L$ with open boundary conditions. Based on the empty interval
approach the time evolution is described by a system of $\frac{L(L+1)}{2}$
differential equations which is solved analytically. An exact expression for
the concentration is derived and its finite-size scaling behaviour is
investigated. The scaling function is found to be
independent of initial conditions. The scaling
function and the correction function for open boundary conditions are
different from those for periodic boundary conditions.\\[3mm]
\rule{5cm}{0.2mm}
\vspace{-4mm}
\begin{flushleft}
\parbox[t]{3.5cm}{\bf Key words:}
\parbox[t]{12.5cm}{Reaction-diffusion systems, finite-size scaling,
             non-equilibrium statistical mechanics, coagulation model}
\\[2mm]
\parbox[t]{3.5cm}{\bf PACS numbers:}
\parbox[t]{12.5cm}{05.40.+j, 05.70.Ln, 82.20.Mj}
\end{flushleft}
\normalsize
\vspace{3cm}
\begin{flushleft}
BONN HE-94-01\\
cond-mat/9402018\\
Bonn University\\
January 1994\\
ISSN-0172-8733
\end{flushleft}
\thispagestyle{empty}
\mbox{}
\newpage
\setcounter{page}{1}
\pagestyle{plain}
%
\section{Introduction}
\hspace{\parindent}
In this article we continue the investigation of the finite-size scaling
properties of reaction-diffusion models defined
on an one-dimensional lattice of $L$ sites.
We consider the example of a special model with
only one kind of particles denoted by $A$ in which the elementary processes
are diffusion, coagulation and decoagulation. This model is integrable
and allows the occupation probability $<\!n_i\!>(t)$ of site~$i$
to be computed analytically. A detailed analysis
for {\it periodic boundary conditions} has been given in our
previous article \cite{Paper1} (hereafter referred to as I),
where we focused our attention on the finite-size scaling properties
of the concentration:
\begin{equation}
c(L,t) \;=\; \frac 1 L \,\sum_{i=1}^L\,<\!n_i\!>(t)
\end{equation}
in the scaling limit $L \rightarrow \infty$, $t \rightarrow \infty$
keeping the scaling variable $z=\frac{4Dt}{L^2}$ fixed. We computed the
corresponding scaling functions exactly and observed that they
do not depend on the initial conditions.

In this paper we consider the same model with {\it open boundary
conditions}. In particular we want to understand how the boundary
conditions modify the finite-size scaling properties of the
concentration $c(L,t)$.

Let us first describe the model and outline the method which allows
us to compute the concentration. The model is defined by
three elementary processes:
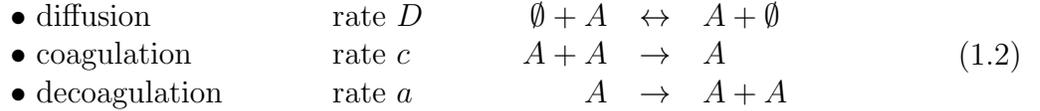
\begin{equation}
\label{Prozesse}
\mbox{
\begin{tabular}{llrcl}
$\bullet$ diffusion &
rate $D$ & $\nil+A$ & $\leftrightarrow$ & $A+\nil$ \\
$\bullet$ coagulation & rate $c$ & $A+A$ & $\rightarrow$ & $A$ \\
$\bullet$ decoagulation \hspace{9mm} & rate $a$
\hspace{9mm} & $A$ & $\rightarrow$ & $A+A$
\end{tabular}}\end{equation}
where $A$ and $\nil\;$ denote an occupied and an empty site, respectively.
The empty interval approach, which we will use to calculate the concentration,
requires that coagulation and diffusion take place with equal rates.
Therefore we have to choose $c=D$. Furthermore it is convenient to fix the
time scale so that $D=1$.
The system is now controlled only by one parameter, namely the
decoagulation rate $a$. Since all rates are chosen to be left-right
symmetric the model is invariant under space reflection.
In paper I we summarized how the master equation of the
reaction-diffusion system defined by (\ref{Prozesse}) can be
mapped onto a one-dimensional quantum chain \cite{Alcaraz}.
In this formulation the time evolution of the model
is described by the euclidean Schr\"odinger equation:
\begin{equation}
\frac{\partial}{\partial t} |P\rangle \;=\;
-\tilde{H} |P \rangle
\end{equation}
where the Hamiltonian $\tilde{H}$ is given by:
\begin{equation}
\label{HamiltonSumme}
\tilde{H} \;=\; \sum_{i=1}^{L-1} \,\tilde{H}_i
\end{equation}
and:
\begin{eqnarray}
\tilde{H}_i &=& +\,2E_i^{11}\otimes E_{i+1}^{11}\,
+\,(1+a)(E_i^{00}\otimes E_{i+1}^{11}+ E_i^{11}\otimes E_{i+1}^{00})\\
&&-\,(E_i^{01}\otimes E_{i+1}^{10}+ E_i^{10}\otimes E_{i+1}^{01}
+ E_i^{01}\otimes E_{i+1}^{11}+ E_i^{11}\otimes E_{i+1}^{01})\nn\\
&&-\,a(E_i^{10}\otimes E_{i+1}^{11}+E_i^{11}\otimes E_{i+1}^{10})\,.\nn
\end{eqnarray}
Here $E_i^{\alpha \beta}$ are $2 \times 2$ matrices defined by
$(E_i^{\alpha,\beta})_{k,l} = \delta^\alpha_k \delta^\beta_l$ acting on the
$i^{\mbox{\scriptsize{th}}}$ site.
Notice that open boundary conditions are already implemented in
Eq. (\ref{HamiltonSumme}). Our aim is to compute the mean value of the
concentration:
\begin{equation}
c(L,t) \;=\; \frac 1 L \, \sum_{i=1}^L\,<\!n_i\!>(t) \;=\;
\frac 1 L \, \sum_{i=1}^L \langle 0 |
E^{11}_i\,\exp(-\tilde{H}t) | P_0 \rangle
\label{konz}
\end{equation}
for an arbitrary initial state $|P_0\rangle$ (for notations
we refer to paper I).

In our first paper we have shown that for
$a \neq 0$ the Hamiltonian $\tilde{H}$ with periodic boundary
conditions can be mapped onto the well-known anisotropic
$XY$-chain with periodic boundary conditions. Since the
corresponding similarity transformation is local, it also
applies to the case of open boundary conditions.
One obtains the Hamiltonian:
\begin{equation}
\label{xy}
H^{XY} \; = \;
-\frac{\eta}{2}\sum_{i=1}^{L-1}\:
\left( \,\eta\:\sigma_{i}^{x}\sigma_{i+1}^{x}
\; + \; \eta^{-1}\sigma_{i}^{y}\sigma_{i+1}^{y}
\; + \; \sigma_{i}^{z} \; + \;
\sigma_{i+1}^{z} \;-\; \eta-\eta^{-1} \right) \;
\end{equation}
where:
\begin{equation}
\eta \;=\; \sqrt{1+a}\,
\label{eta}
\end{equation}
is a real parameter. Notice that although the original system
(\ref{HamiltonSumme}) is defined with open boundary conditions,
in Eq. (\ref{xy}) we do {\it not}
obtain a quantum chain with free boundaries since there are additional fields
$\frac{2}{\eta}(\sigma_1^z+\sigma_L^z)$ at the ends of the chain.
The $XY$-chain is integrable, i.e. it
can be diagonalized exactly in terms of free fermions,
which allows to write the Hamiltonian (\ref{xy}) in the diagonal
form \cite{LSM}:
\begin{equation}
\label{diagonal}
H^{XY} \;=\; \sum_{k=0}^{L-1} \, \Lambda_k\,
a_k^\dagger a_k\,,
\end{equation}
where $a_k^\dagger$ and $a_k$ are fermionic operators:
\begin{equation}
\{a_k^\dagger,a_l\} \;=\; \delta_{k,l}\,.
\end{equation}
The $\Lambda_k$ are the corresponding
fermionic excitation energies and are given by:
\begin{eqnarray}
\label{ZeroMode}
\Lambda_0 &=& 0\;, \\
\label{Dispersion}
\Lambda_k &=& \eta \, \biggl(\eta+\eta^{-1}-2\cos \frac{\pi k}{L}\biggr)\;,
\hspace{1cm} k=1,\ldots,L-1\;.
\label{ewert}
\end{eqnarray}
The excitation with vanishing energy (\ref{ZeroMode})
for arbitrary $\eta$
is related to a hidden quantum group symmetry \cite{HayeVlad}
and implies the levels of the spectrum to be at least
two-fold degenerated. The second equation (\ref{Dispersion})
is just the dispersion relation of the $XY$-chain and we observe that
the system is massless for $\eta=1$ ($a=0$).
Taking all combinations of the excitation
energies $\L_k$ into account, one can construct the spectrum
of $H^{XY}$ and therewith the spectrum of $\tilde{H}$.

One way to compute the concentration $c(L,t)$ would be to
diagonalize $\tilde{H}$ directly. This turns out to be very
complicated since it is not known in advance which of the eigenvectors
enter the expression for the concentration and thus all of them have to be
calculated.
Therefore we prefer a different approach
based on so-called empty intervals probabilities
instead of spin configurations \cite{Avraham,Doering}
(for a compact summary we refer to paper I, the
most general formulation of this approach
can be found in \cite{Peschel}).
The time evolution
of the probability $\Omega(j,n,t)$ of
finding an empty interval of length $n$
extending from site $j+1-\frac n 2$ to site $j+\frac n 2 $ is described
by a closed system of linear
differential equations.
Since $1-\O(j-\half,1,t)$ is just the probability to find the
$j^{\mbox{\scriptsize th}}$ site occupied
the concentration $c(L,t)$ is simply given by:
\begin{equation}
\label{concentration}
c(L,t) \;=\; \frac 1 L \, \sum_{j=1}^L \, <\!n_j\!>(t) \;=\;
\frac 1 L \sum_{j=1}^L \, \Bigl(1-\Omega(j-\frac12,1,t)\Bigr)\,.
\end{equation}
A detailed derivation of this formalism can be found in paper I. Here
we present a modified version of the differential equations for
open boundary conditions. If the interval
does not touch the boundaries we have:
\begin{eqnarray}
\label{eq1}
\frac{\partial\Omega(j,n,t)}{\partial t } &=&
\e^2\, \Bigl( \Omega(j-\half,n+1,t) + \Omega(j+\frac12,n+1,t)\Bigr) \\
&&+ \;\Omega(j-\frac12,n-1,t) + \Omega(j+\frac12,n-1,t)
- 2\,(\e^2+1)\,\Omega(j,n,t) \nonumber\,,
\end{eqnarray}
where $\frac{n}{2}< j <L-\frac{n}{2}$ and $n=1,\cdots,L-1$.
If on the other hand the interval touches the boundaries one
obtains the `boundary equations':
\begin{eqnarray}
\label{eq2}
\frac{\partial\Omega(\frac{n}{2},n,t)}{\partial t} &=&
\e^2\Omega(\frac{n+1}{2},n+1,t) \\
&&+ \, \Omega(\frac{n-1}{2},n-1,t)
-(\e^2+1)\,\Omega(\frac{n}{2},n,t)\,, \nonumber \\
\label{eq3}
\frac{\partial\Omega(L-\frac{n}{2},n,t)}{\partial t} &=&
\e^2\Omega(L-\frac{n+1}{2},n+1,t) \\
&&+ \, \Omega(L-\frac{n-1}{2},n-1,t)
-(\e^2+1)\,\Omega(L-\frac{n}{2},n,t)\,,\nonumber\\
\label{eq4}
\frac{\partial\Omega(\frac{L}{2},L,t)}{\partial t}  &=& 0\,.
\end{eqnarray}
As in paper I in all equations we use the convention:
\begin{equation}
\label{convention}
\Omega(j,0,t)=1\,.
\end{equation}
Therefore this system of equations
is inhomogeneous. A particular solution is obviously given by:
\begin{equation}
\label{particular}
\Omega(j,n,t)\;=\; 1
\end{equation}
corresponding to an empty lattice. In this notation the space reflection
is described by the map:
\be
P\;:\;(j,n)\mapsto(L-j,n)\;.
\ee
As one can
easily check the system of differential equations (\ref{eq1})-(\ref{eq4})
is invariant under~$P$. Notice that, in
opposition to the case of periodic boundary conditions where the Fourier
transformed system has only $L-1$ degrees of freedom,
this system has $L(L+1)/2$ degrees of freedom.
It grows quadratically with lattice length $L$ and thus it is more
difficult to solve.
The central mathematical aspect of this paper is
the exact solution of this problem.

Because of the reduced number of degrees of freedom compared to
the original physical system the set of differential equations
(\ref{eq1})-(\ref{eq4}) can only cover a part of the full dynamics
of the Hamiltonian (\ref{HamiltonSumme}). As we will see in Sec.
\ref{sec:solution} this
part corresponds just to those states of the $XY$ chain
with only one or two excited fermions.
The advantage of the empty interval formalism is that $c(L,t)$
can be computed easily via Eq. (\ref{concentration}).

This article is organized as follows. In Sec. \ref{sec:solution}
we solve the system
of differential equations (\ref{eq1})-(\ref{eq4}). We then
derive exact expressions for $c(L,t)$ for special initial
conditions in Sec.
\ref{sec:fss} and analyse the corresponding finite-size scaling
properties. Finally we summarize our results.
%
\section{Solution of the System of Differential Equations}
\label{sec:solution}
\hspace{\parindent}
In this section we solve the system of differential equations
(\ref{eq1})-(\ref{eq4}). For this purpose we first
determine the complete set of solutions
$\phi_\L(j,n,t)$
of the homogeneous eigenvalue problem:
\begin{equation}
\label{evp}
\frac{\partial}{\partial t} \phi_\L(j,n,t) \;=\;
- \L\, \phi_\L(j,n,t)\,,
\end{equation}
where Eq. (\ref{convention}) has to be replaced by:
\begin{equation}
\label{homogen}
\phi_\L(j,0,t)=0
\end{equation}
in order to impose homogeneous boundary conditions.
It will be convenient to introduce the
notations:
\begin{equation}
x=j+\frac{n}{2}\,,\hspace{2cm} y=j-\frac{n}{2}
\end{equation}
where $x$ and $y$ take the values $0\leq y\leq x\leq L$.
In this notation the space reflection reads:
\be
\label{reflex}
P\;:\;(x,y) \mapsto (L-y,L-x)\;.
\ee
For $0<y<x<L$ the system of differential equations (\ref{eq1})-(\ref{eq4})
now has the form:
\ba
\label{eq5}
\frac{\partial\phi(x,y,t)}{\partial t } &=&
\e^2\, \Bigl( \phi(x+1,y,t) + \phi(x,y-1,t)\Bigr) \\
&&+ \;\phi(x-1,y,t) + \phi(x,y+1,t)
- 2(\e^2+1)\,\phi(x,y,t) \nn\,,
\ea
\ba
\label{eq6}
\frac{\partial\phi(x,0,t)}{\partial t} &=&
\e^2 \phi(x+1,0,t) + \, \phi(x-1,0,t)
-(\e^2+1)\,\phi(x,0,t)\;,\\
\label{eq7}
\frac{\partial\phi(L,y,t)}{\partial t} &=&
\e^2 \phi(L,y-1,t) + \, \phi(L,y+1,t)
-(\e^2+1)\,\phi(L,y,t)\;,\\
\label{eq8}
\frac{\partial\phi(L,0,t)}{\partial t}  &=& 0\;.
\ea
For $0\leq x=y\leq L$ we have:
\be
\label{hom}
\phi(x,x,t)=0\;,
\ee
or:
\be
\label{inhom}
\phi(x,x,t)=1
\ee
if we choose the system to be homogeneous or inhomogeneous, respectively.
For easy reference we illustrated
the organization of the complete system for $L=6$
sites in Fig. 1. Here the $L(L+1)/2$ boxes represent the degrees of
freedom $\p(x,y)$. The number in each box denotes the strength of
the diagonal contribution in the differential equations, where
$d=-(2+a)=-(1+\e^2)$, while the arrows denote the flow of probability
to the neighbours with respect to the couplings
\begin{eqnarray}
\rightarrow,\downarrow: &\hspace{5mm}& 1 \nonumber \\
\leftarrow,\uparrow: &\hspace{5mm}& 1+a=\e^2\,. \nonumber
\end{eqnarray}
\begin{figure}[0mm]
%
%
%
\def\bu{\begin{picture}(2,2)
        \put(-0.88,-0.28){\makebox[8mm][c]{\rule[-1.5mm]{0mm}{6mm}0}}
        \end{picture}}
%
%
\def\ba{\begin{picture}(2,2)
        \put(-0.88,-0.28){\framebox[8mm][c]{\rule[-1.5mm]{0mm}{6mm}0}}
        \end{picture}}
%
%
\def\bb{\begin{picture}(2,2)
        \put(-0.88,-0.28){\framebox[8mm][c]{\rule[-1.5mm]{0mm}{6mm}d}}
        \end{picture}}
%
%
\def\bc{\begin{picture}(2,2)
        \put(-0.88,-0.28){\framebox[8mm][c]{\rule[-1.5mm]{0mm}{6mm}2d}}
        \end{picture}}
%
%
\def\au{\begin{picture}(2,2)
        \put(0,-0.5){\vector(0,1){1}}
	\end{picture}}
%
%
\def\ad{\begin{picture}(2,2)
        \put(0,0.5){\vector(0,-1){1}}
	\end{picture}}
%
%
\def\aud{\begin{picture}(2,2)
        \put(0.2,0.5){\vector(0,-1){1}}
        \put(-0.2,-0.5){\vector(0,1){1}}
	\end{picture}}
%
%
\def\ar{\begin{picture}(2,2)
        \put(-0.5,0){\vector(1,0){1}}
	\end{picture}}
%
%
\def\al{\begin{picture}(2,2)
        \put(0.5,0){\vector(-1,0){1}}
	\end{picture}}
%
%
\def\arl{\begin{picture}(2,2)
        \put(-0.5,-0.2){\vector(1,0){1}}
        \put(0.5,0.2){\vector(-1,0){1}}
	\end{picture}}
\setlength{\unitlength}{4mm}
\begin{picture}(34,34)
\put (4,4)  {\vector(0,1){28}}
\put (4,4)  {\vector(1,0){28}}
\put (33,3.6) {x}
\put (3.8,32.6) {y}
\put (4,6) {\line(-1,0){0.2}}
\put (4,10) {\line(-1,0){0.2}}
\put (4,14) {\line(-1,0){0.2}}
\put (4,18) {\line(-1,0){0.2}}
\put (4,22) {\line(-1,0){0.2}}
\put (4,26) {\line(-1,0){0.2}}
\put (4,30) {\line(-1,0){0.2}}
\put (6,4) {\line(0,-1){0.2}}
\put (10,4) {\line(0,-1){0.2}}
\put (14,4) {\line(0,-1){0.2}}
\put (18,4) {\line(0,-1){0.2}}
\put (22,4) {\line(0,-1){0.2}}
\put (26,4) {\line(0,-1){0.2}}
\put (30,4) {\line(0,-1){0.2}}
\put (3,5.7) {0}
\put (3,9.7) {1}
\put (3,13.7) {2}
\put (3,17.7) {3}
\put (3,21.7) {4}
\put (3,25.7) {5}
\put (3,29.7) {6}
\put (5.8,2.6) {0}
\put (9.8,2.6) {1}
\put (13.8,2.6) {2}
\put (17.8,2.6) {3}
\put (21.8,2.6) {4}
\put (25.8,2.6) {5}
\put (29.8,2.6) {6}
\put (6,6)  {\bu}
\put (8,6)  {\ar}
\put (10,6) {\bb}
\put (12,6) {\arl}
\put (14,6) {\bb}
\put (16,6) {\arl}
\put (18,6) {\bb}
\put (20,6) {\arl}
\put (22,6) {\bb}
\put (24,6) {\arl}
\put (26,6) {\bb}
\put (28,6) {\al}
\put (30,6) {\ba}
\put (14,8) {\au}
\put (18,8) {\au}
\put (22,8) {\au}
\put (26,8) {\au}
\put (30,8) {\au}
\put (10,10) {\bu}
\put (12,10) {\ar}
\put (14,10) {\bc}
\put (16,10) {\arl}
\put (18,10) {\bc}
\put (20,10) {\arl}
\put (22,10) {\bc}
\put (24,10) {\arl}
\put (26,10) {\bc}
\put (28,10) {\al}
\put (30,10) {\bb}
\put (14,12) {\ad}
\put (18,12) {\aud}
\put (22,12) {\aud}
\put (26,12) {\aud}
\put (30,12) {\aud}
\put (14,14) {\bu}
\put (16,14) {\ar}
\put (18,14) {\bc}
\put (20,14) {\arl}
\put (22,14) {\bc}
\put (24,14) {\arl}
\put (26,14) {\bc}
\put (28,14) {\al}
\put (30,14) {\bb}
\put (18,16) {\ad}
\put (22,16) {\aud}
\put (26,16) {\aud}
\put (30,16) {\aud}
\put (18,18) {\bu}
\put (20,18) {\ar}
\put (22,18) {\bc}
\put (24,18) {\arl}
\put (26,18) {\bc}
\put (28,18) {\al}
\put (30,18) {\bb}
\put (22,20) {\ad}
\put (26,20) {\aud}
\put (30,20) {\aud}
\put (22,22) {\bu}
\put (24,22) {\ar}
\put (26,22) {\bc}
\put (28,22) {\al}
\put (30,22) {\bb}
\put (26,24) {\ad}
\put (30,24) {\aud}
\put (26,26) {\bu}
\put (30,26) {\bb}
\put (30,28) {\ad}
\put (30,30) {\bu}
\end{picture}
\caption{Structure of the system of
differential equations for $L=6$ sites.
The figure is explained in the text.}
\end{figure}
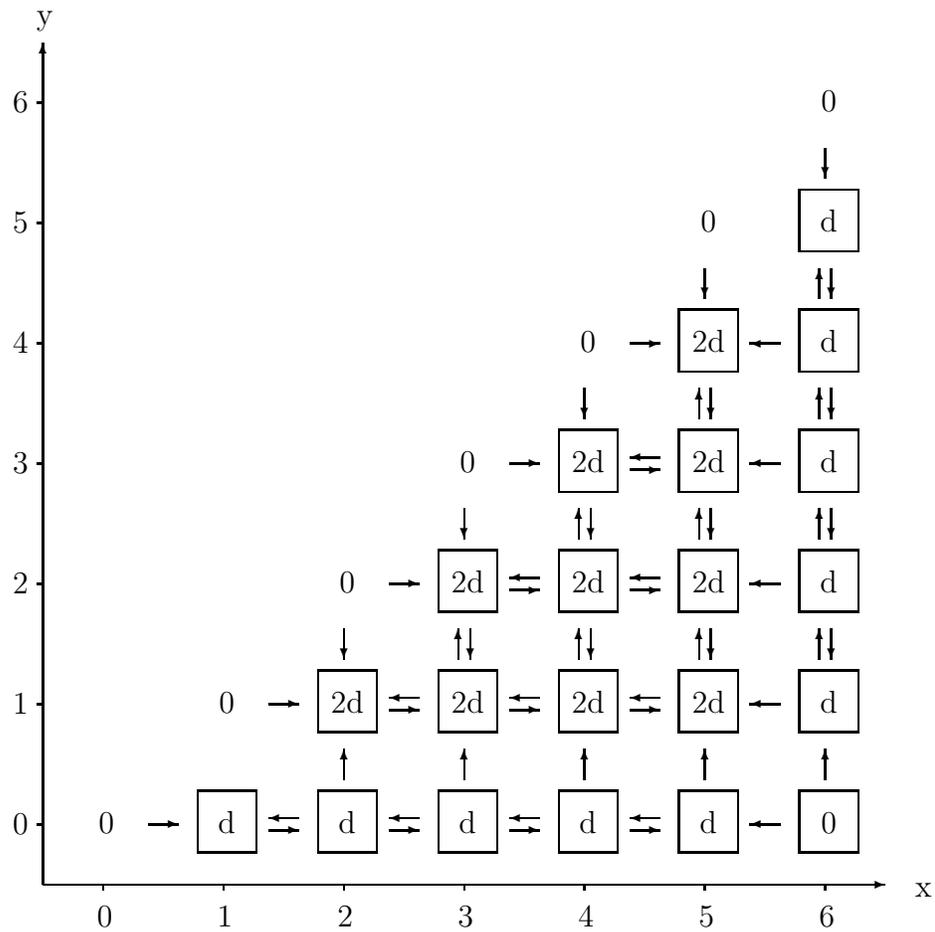
The zeros on the hypotenuse $x=y$ represent the contributions from the
inhomogenous part of the system. Since we consider only the homogenous
equation due to Eq. (\ref{hom}) these terms have to be neglegted
throughout this section.

Before calculating the solutions of the eigenvalue problem
of Eq. (\ref{evp}) we give a short
description how to proceed.
Looking at Fig. 1 we find that the boxes on the short sides of the triangle
form a decoupled $(2L-1)$-dimensional subsystem of differential equations.
As a consequence we expect two kinds of solutions:
\begin{itemize}
\item
The first kind is obtained
by solving the subsystem separately.
Since each solution imposes inhomogeneous Dirichlet boundary conditions
it can be extended uniquely to the interior of the triangle.
This procedure yields a set of $2L-1$ solutions of Eq. (\ref{evp})
forming a complete set of eigenfunctions for the short sides of the triangle.
\item
The second kind of solutions is now calculated using the completeness
of the first kind solutions. They are forced to vanish on the boundary
and therefore to obey homogeneous Dirichlet boundary conditions.
\end{itemize}
Let us now study the solution of (\ref{evp}) in more detail.
For the moment we choose $\e=1$ (see Eq.~(\ref{eta})).
Then the differential equations acting on
the short sides of the triangle are:
\begin{eqnarray}
D_x^2\,\p_\L(x,0) &=& -\L\,\p_\L(x,0), \label{tria1}\\
D_y^2\,\p_\L(L,y) &=& -\L\,\p_\L(L,y) \label{tria2}\,,
\end{eqnarray}
where $x,y=1,\cdots,L-1$. Here $D_x^2$ and $D_y^2$ are discrete second
derivative operators in $x$ and $y$ direction, respectively. $D_x^2$ for
example is defined by
$D_x^2\phi(x,y)=\phi(x+1,y)+\phi(x-1,y)-2\phi(x,y)$.

The probability for an empty lattice
$\p_\L(L,0)$ decouples completely and remains constant in time.
Thus $\p_\L(L,0)$ takes nonzero values only if $\L=0$. In this case the second
derivative of $\p_{\L=0}$ vanishes and $\p_0$ becomes a linear function of
$x$ and $y$, namely $\p_0(x,y)\sim x-y$ due to Eq. (\ref{hom}).
For $\L\neq0$ the condition $\p_\L(L,0)=0$ gives an additional constraint which
determines the solutions of Eqs. (\ref{tria1}) and (\ref{tria2})
to be simple oscillations:
\begin{eqnarray}
\label{solution}
\p(x,0) \sim \sin \frac{\pi k x}{L}\;,\hspace{5mm}
\p(L,y) \sim \sin \frac{\pi k y}{L}\;,\\[4mm]
\L_k = 2\,\biggl(1-\cos\frac{\pi k}{L}\biggr)\nn\,,
\end{eqnarray}
where $k=1,\ldots,L-1$.
The values of $\L$ give the spectrum of eigenvalues.
Comparing it with the spectrum of the $XY$-chain (\ref{diagonal})
we recognize that it corresponds to one-fermion excitations.
For a given $\L_k\neq0$ every linear combination of the solutions of Eq.
(\ref{solution}) solves Eqs. (\ref{tria1}) and (\ref{tria2}).
Thus these eigenvalues are twofold degenerated. However, we can choose the
functions $\p_k$ to be eigenfunctions of the space reflection $P$
(see (\ref{reflex}), too.
It will be convenient to introduce the notation
$P\p_k^{\pm}=\pm(-1)^k\p_k^\pm$. So far we have determined the required $2L-1$
solutions on the short sides of the triangle which still have to be extended
to the interior.
Here the differential equations act according to $(D_x^2+D_y^2)\p=-\L\p$
where the operator on the left side is just the discrete Laplacian.
The solutions are given below in Eqs. (\ref{sol2}) and (\ref{sol2a}).

Now we turn to the calculation of the remaining  $\frac{(L-1)(L-2)}{2}$
solutions of the interior of the triangle. These solutions
are plane waves (\ref{sol3}) vanishing on the boundary of the triangle.
The corresponding eigenvalues $\L=\L_k+\L_l$ are identified as two-fermion
excitations of the $XY$-chain.

Summing up all results we have three types of solutions:
\begin{itemize}
\item The stationary solution:
\begin{equation}
\label{sol1}
\p_0(x,y) \;=\;\frac{x-y}{L}\;.
\end{equation}
\item One-fermion excitations $(k=1,\ldots,L-1)$:
\begin{eqnarray}
\label{sol2}
\p_k^+(x,y) &=& \frac{1}{\sqrt{L}}\biggl(
     \sin\frac{\pi k}{L}x-\sin\frac{\pi k}{L}y\biggr)\;, \\
\label{sol2a}
\p_k^-(x,y) &=& \frac{1}{\sqrt{L}}\biggl((1-\frac{2y}{L})\sin\frac{\pi k}{L}x-
                   (1-\frac{2x}{L})\sin\frac{\pi k}{L}y\biggr)\;.
\end{eqnarray}
\item Two-fermion excitations $(k,l=1,\ldots,L-1; k < l)$:
\begin{equation}
\label{sol3}
\Phi_{kl}(x,y) \;=\; \frac{2}{L}\biggl(
     \sin\frac{\pi k}{L}x \, \sin\frac{\pi l}{L}y
   - \sin\frac{\pi l}{L}x \, \sin\frac{\pi k}{L}y\biggr)\;.
\end{equation}
\end{itemize}
This is the complete set of $L(L+1)/2$ solutions of the homogeneous
system. For $\e \neq 1$ we obtain deformations of these solutions.
In order to avoid too much details, we only want
to present our results:
\begin{itemize}
\item The stationary solution:
\begin{equation}
\label{sol4}
\p_0(x,y) \;=\;\frac{1-\e^{2(y-x)}}{1-\e^{-2L}}
\end{equation}
with eigenvalue $\L=0$ and parity $P=+1$.
\item One-fermion excitations $(k=1,\ldots,L-1)$:
\begin{equation}
\label{sol5}
\p_{k}^\pm(x,y)\;=\;\frac{1}{\sqrt{L}}\frac{\e^{y-x}}{1\pm\e^L}\biggl(
(\e^y\pm\e^{L-y})\sin\frac{\pi k}{L}x
- (\e^x\pm\e^{L-x})\sin\frac{\pi k}{L}y\biggr)
\end{equation}
with eigenvalue $\L_k=\e(\e+\e^{-1}-2\cos{\frac{\pi k}{L}})$
(see Eq. (\ref{ewert})) and parity $P=\pm(-1)^k$.
\item Two-fermion excitations $(k,l=1,\ldots,L-1; k < l)$:
\begin{equation}
\label{sol6}
\Phi_{kl}(x,y)\;=\;\frac{2}{L}\e^{y-x}\biggl(
     \sin\frac{\pi k }{L}x \, \sin\frac{\pi l }{L}y
   - \sin\frac{\pi l }{L}x \, \sin\frac{\pi k }{L}y\biggr)
\end{equation}
with eigenvalues
$\L_{kl}=\e(2\e+2\e^{-1}-2\cos{\frac{\pi k}{L}}-2\cos{\frac{\pi l}{L}})$
and parity $P=(-1)^{k+l+1}$.
\end{itemize}
It is easy to check that these solutions reduce to Eqs.
(\ref{sol1})-(\ref{sol3}) in the limit $\e \rightarrow 1$.

Let us now study the orthogonality relations of the different types of
solutions.
The functions $\p_k^\pm,\; k\neq 0$
form a complete system
of eigenfunctions on the short sides of the triangle
except the point $(L,0)$. The corresponding
scalar product denoted by
$\randsp{}{}$
takes only these points into account.
For arbitrary functions $f,g$ we define:
\be
\randsp{f}{g}=\sum_{x=1}^{L-1}\e^{2x}f(x,0)g(x,0)+
           \sum_{y=1}^{L-1}\e^{2(L-y)}f(L,y)g(L,y)\;.
\label{SP1}
\ee
The solutions (\ref{sol5}) are normalized in such a way that:
\be
\randsp{\p_k^\a}{\p_{k'}^\b}=\d_{kk'}\d_{\a\b}
\label{orth1}
\ee
where $k,k'=1,\ldots,L-1$ and $\a,\b\;=\;\pm$.
For the functions $\Phi_{kl}$ (\ref{sol6}) we define a scalar product in the
interior of the triangle:
\be
\bulksp{f}{g}
=\sum_{x=1}^{L-1}\sum_{y=1}^{x-1}\e^{2(x-y)}f(x,y)g(x,y)\,.
\label{SP2}
\ee
Then the functions $\Phi_{kl}$ (\ref{sol6}) obey the relation:
\be
\bulksp{\Phi_{kl}}{\Phi_{k'l'}}=\d_{kk'}\d_{ll'}\,.
\label{orth2}
\ee
These relations are independent of the value of $\e$.
Unfortunately the solutions $\p_k^\pm$ take nonzero values in the interior
of the triangle. Therefore the scalar product
$\bulksp{\Phi_{kl}}{\p_{k'}^\pm}$
is not equal to zero. These projections have to be taken into account if
one expands an arbitrary function $\O(x,y)$ in terms of
the eigenfunctions as we will do in the next section.
%
\section{Finite-Size Scaling of the Concentration}
\label{sec:fss}
\hspace{\parindent}
In this section we use the complete set of eigenfunctions
found in Sec. \ref{sec:solution} in order to calculate the time evolution
of the concentration for an initially fully occupied lattice and
an arbitrary value of $\e$. In the case of $\e=1$ the finite-size scaling
limit is computed and compared with the corresponding result
for periodic boundary conditions.

Before calculating the concentration for special initial conditions
we consider the general case. Let $\O_0(x,y)$ be an arbitrary initial
choice of empty interval probabilities satisfying condition (\ref{convention}).
The time evolution then takes the form:
\be
\O(x,y,t)=1+\w_0\p_0(x,y)+\sum_{k=1}^{L-1}\sum_{\a=\pm}\w_k^\a\p_k^\a(x,y)
e^{-\L_kt}+\sum_{k=1}^{L-2}\sum_{l=k+1}^{L-1}\w_{kl}\Phi_{kl}(x,y)
e^{-(\L_k+\L_l)t}
\label{eq:timeev}
\ee
where the constant 1-function is just the particular solution
(\ref{particular}) and $\L_k$ is defined in Eq. (\ref{ewert}).
The coefficients $\w_0,\;\w_k^\pm$ and $\w_{kl}$ are obtained by
expanding $\O_0(x,y)$ in terms of the eigenfunctions
(\ref{sol4})-(\ref{sol6}).
Therefore we first consider Eq. (\ref{eq:timeev}) for $t=0$ at
the point $(L,0)$.
Since all eigenfunctions except $\p_0$ vanish at this point we have:
\be
\w_0=\O_0(L,0)-1\;.
\ee
Then we consider the short sides of the triangle, where the eigenfunctions
$\Phi_{kl}$ vanish. Here the coefficients $\w^\pm_k$ can be obtained
using the relation (\ref{orth1}):
\be
\label{om1}
\w_k^\pm = \randsp{\p_k^\pm}{\O_0-1-\w_0\p_0}\;.
\ee
Finally we calculate the remaining coefficients $\w_{kl}$ with the help of
Eq. (\ref{orth2}):
\be
\label{om2}
\w_{kl}=\bulksp{\Phi_{kl}}{\O_0-1-\w_0\p_0}
-\sum_{k'=1}^{L-1}\sum_{\a=\pm}\w_{k'}^\a\bulksp{\Phi_{kl}}{\p_{k'}^\a}\;.
\ee
We now apply this general procedure to the case of an uncorrelated initial
state where the occupation probability $p$ is equal for all sites (see
paper I). Then we have:
\be
\O_0(x,y)=(1-p)^{x-y}\;.
\ee
Straightforward but rather laborious calculations show that
the concentration is given by:
\ba
c(t,L)&=&\frac{1-\e^{-2}}{1-\e^{-2L}}(1-(1-p)^L) \\&& +\, \nonumber
\frac{4\,\,(1-\e^{-2})}{L^2(\e^L-\e^{-L})}\,\sum_{k=1}^{L-1}\,
(1-(1-p)^L\eta^L(-1)^k)(\e^L+\e^{-L}-2(-1)^k) \\ && \hspace{38mm} \times
\frac{\sin^2(\frac{\pi k}{L})}{\L_k} \nonumber
\biggl( \frac{1}{\L_k}-\frac{1}{\tilde{\L}_k} \biggr) \exp{(-\L_kt)}\\
&&+\,\frac{8}{L^3 \eta}\sum_{k=1}^{L-1}
\sum_{\scriptstyle l=k+1\atop\scriptstyle k+l\;\mbox{\scriptsize odd}}^{L-1}
\frac{\sin^2\frac{\pi k}{L}\;\sin^2\frac{\pi l}{L}}
             {\cos\frac{\pi k}{L}-\cos\frac{\pi l}{L}} \nonumber
\Biggl[ \frac{1}{\cos\frac{\pi k}{L}-\cos\frac{\pi l}{L}}
        \biggl( \frac{1}{\L_k}+\frac{1}{\L_l}
               -\frac{1}{\tilde{\L}_k}-\frac{1}{\tilde{\L}_l} \biggr) \\
\nonumber && \hspace{58mm} + \,
2 \biggl( \frac{1}{\L_k}-\frac{1}{\tilde{\L}_k} \biggr)
\biggl( \frac{1}{\L_l}-\frac{1}{\tilde{\L}_l} \biggr)
\,\eta^L (1-p)^L (-1)^k \\ \nonumber && \hspace{58mm} -
\frac{2}{\L_k\tilde{\L}_l} + \frac{2}{\L_l\tilde{\L}_k}\Biggr]\,\,
\exp{(-(\L_k+\L_l)t)}\;, \nonumber
\ea
where $\tilde{\L}_k$ is defined by:
\begin{equation}
\tilde{\L}_k \;=\;
\e\biggl((1-p)\eta +(1-p)^{-1} \eta^{-1} - 2 \cos \frac{\pi k}{L}\biggr)\;.
\end{equation}
Now we consider the finite-size scaling limit of the concentration.
As pointed out in paper I we assume that in
the finite-size scaling regime $L\rightarrow\inf$,$\,t\rightarrow\inf$,
$z=\frac{4t}{L^2}$ fixed the concentration behaves like:
\be
c(z,L)=L^x[F_0(z)+L^{-y}F(z)+\cdots]\;.
\label{eq:fss}
\ee
In this formula $F_0(z)$ denotes the scaling function and $x$ the
scaling exponent. $L^yF(z)$ is the leading correction term. The
functions $F_0(z)$ and $F(z)$ are supposed to depend only on $z$.
Since the finite-size scaling hypothesis is only valid for critical systems
corresponding to massless quantum chains we have to restrict
ourselves to the case $\e=1$.
In this case the concentration takes the form:
\ba
c(t,L)&=&\frac{1-(1-p)^L}{L} \\
&&+\,\frac{4}{L^3}\sum_{k=0}^{L-1}
\sum_{\scriptstyle l=1\atop\scriptstyle k+l\;\mbox{\scriptsize odd}}^{L-1}
\frac{\sin^2\frac{\pi k}{L}\;\sin^2\frac{\pi l}{L}}
             {\cos\frac{\pi k}{L}-\cos\frac{\pi l}{L}} \nonumber
\Biggl[ \frac{1}{\cos\frac{\pi k}{L}-\cos\frac{\pi l}{L}}
        \biggl( \frac{1}{\L_k}+\frac{1}{\L_l}
               -\frac{1}{\tilde{\L}_k}-\frac{1}{\tilde{\L}_l} \biggr) \\
\nonumber && \hspace{58mm} + \,
2 \biggl( \frac{1}{\L_k}-\frac{1}{\tilde{\L}_k} \biggr)
\biggl( \frac{1}{\L_l}-\frac{1}{\tilde{\L}_l} \biggr)
(1-p)^L (-1)^k \\ \nonumber && \hspace{58mm} -
\frac{2}{\L_k\tilde{\L}_l} + \frac{2}{\L_l\tilde{\L}_k}\Biggr]\,\,
\exp{(-(\L_k+\L_l)t)}\;. \nonumber
\ea
Performing now the finite-size scaling limit we get for the functions
$F_0(z)$ and $F(z)$:
\ba
\label{scalfn}
F_0(z)& = & 1+\frac{16}{\pi^2}\sum_{k=0}^{\inf}
\sum_{\scriptstyle l=1\atop\scriptstyle k+l\;\mbox{\scriptsize odd}}^{\inf}
              \frac{k^2+l^2}{(k^2-l^2)^2}
              \exp\biggl(-\frac{z\pi^2}{4}(k^2+l^2)\biggr)\;,\\[4mm]
\label{corrfn}
F(z) &=& \sum_{k=0}^{\inf}
\sum_{\scriptstyle l=1\atop\scriptstyle k+l\;\mbox{\scriptsize odd}}^{\inf}
     \biggl(\frac{z}{3}\pi^2\frac{(k^2+l^2)(k^4+l^4)}{(k^2-l^2)^2}-
     \frac{8}{3}\frac{(1+6\frac{1-p}{p^2})(k^4+l^4)+k^2l^2}{(k^2-l^2)^2}\biggr)
\\ && \hspace{20mm}\times\exp\biggl(-\frac{z\pi^2}{4}(k^2+l^2)\biggr)\nn
\ea
and for the exponents $x,\;y$:
\be
x=-1,\hspace{1cm}y=2\;.
\ee
The functions $F_0$ and $F$ can be written in a simpler form by
rewriting the sums in Eqs. (\ref{scalfn}) and (\ref{corrfn}) in terms of
$\mu=\frac{k+l-1}{2}$ and $\nu=\frac{k-l-1}{2}$:
\ba
F_0(z) &=& 1-\half g_0(z)\;g_1(z)\;,\\
F(z) &=& -\frac{7z}{12}g_1(z)g_2(z)-\frac{z}{12}g_0(z)g_3(z)-\frac{1}{12}
\biggl(g_0(z)g_2(z)-\biggl(g_1(z)\biggr)^2\biggr)\nn\\
&&-(1+6\frac{1-p}{p^2})\biggl(\frac{1}{6}g_0(z)g_2(z)
+\half\biggl(g_1(z)\biggr)^2\biggr)
\ea
where the functions $g_i(z)$ are defined by:
\ba
g_0(z) &=& \sum_{\mu=-\inf}^{\inf} \frac{-2}{\pi^2 (\mu+\half)^2}
\exp{\biggl(-\frac{z\pi^2}{2} (\mu+\half)^2\biggr)}\;,\\
g_i(z) &=& \frac{\partial^i}{\partial z^i} g_0(z)\;,\;
i=1,\cdots,3\;.
\ea
Especially we have the identity:
\be
g_1(z)=\theta_2\biggl(0,\frac{i\pi z}{2}\biggr)
\ee
where $\theta_2(u,\t)$ is a Jacobi-Theta function.
Now we are able to compare the finite-size scaling behaviour for open and
periodic boundary conditions. In both cases we find the scaling exponent
$x=-1$ and the correction exponent $y=2$.
Thus the critical exponents are not influenced by the boundary conditions.
The scaling function and the correction function on the other hand are
different as can be seen in Fig. 2 and Fig. 3. However we find that
in both cases the scaling functions and the
correction functions are related to Jacobi-Theta functions.
As in the case of periodic boundary conditions we observe that the scaling
function (\ref{scalfn}) is independent of the initial probability
$p$ whereas the correction function (\ref{corrfn}) is not.
%
\section{Conclusions}
\hspace{\parindent}
Based on the empty interval probability approach we solved the
coagulation-decoagulation model defined on a lattice of $L$ sites
with open boundary conditions. In contrast to ordinary mean-field
calculations our results take the microscopic fluctuations of the
system into account. We computed the concentration
$c(L,t)$ exactly and investigated its finite-size scaling properties.
Our aim was to compare the results with those of our previous paper
\cite{Paper1}, where periodic boundary conditions have been
considered.

Mathematically the time evolution of the empty interval probabilities
is described by a system of $L(L+1)/2$ linear differential equations
(in opposition to only $L-1$ equations in the case of periodic
boundary conditions). It is the first time that such a system has
been solved in the theory of reaction-diffusion processes, and we
sketched the general method
how to find the solutions. We obtained
three types of solutions wich can be interpreted as states
with none, one and two fermions in an excited state.

We used the expression for $c(L,t)$ to compute the finite-size scaling
expansion of the concentration and compared the result with the
case of periodic boundary conditions. It turned out that one obtains
the same type of expansion (with the same powers in $1/L$). As in
the periodic case we found that the scaling function is
independend of uncorrelated homogeneous initial conditions.
However, they are
different for open and periodic boundary conditions.
On the other hand the correction function is influenced by the initial
condition similar to the periodic case.
In this paper we only investigated homogeneous uncorrelated initial conditions.
Initial states with small clusters are studied numerically. The results
are presented in \cite{Paper3}. It turns out that the scaling function
remains unchanged even for that kind of initial conditions.

Using the empty interval approach we can compute the local concentration
$c_j(t)$. As mentioned in the introduction, this formalism does not
cover the full dynamical space of the physical system. In particular
it is impossible to compute correlations between concentrations
located at distinct points.
For this reason we want to outline a possible generalization.
An empty interval can be understood as
a two-point object. Therefore it is natural to consider also generalized
$2n$-point objects consisting of $n$ separated empty intervals. These
objects are also expected to be described by a set of linear differential
equations as well. In particular they are nothing but
$n$-point correlation functions of the occupation number if we choose the
lengths of all intervals to be equal to one
(similar to Eq. (\ref{concentration})). Of course we expect a higher
number of fermionic excitations to be involved.
%

\newcounter{fig_count}
\noindent {\Large \bf List of Figures}
\begin{list}
      {Fig. \arabic{fig_count}:}
      {\usecounter{fig_count}  \setlength{\leftmargin}{1.5cm}
               \setlength{\labelsep}{2mm}
               \setlength{\labelwidth}{1.3cm}}
\item Structure of the system of differential equations for
$L\;=\;6$ sites. The figure is explained in the text.
\item Scaling functions for open and
periodic boundary conditions.
\item Correction functions for open and
periodic boundary conditions for initial occupation probability $p=1$.
\end{list}
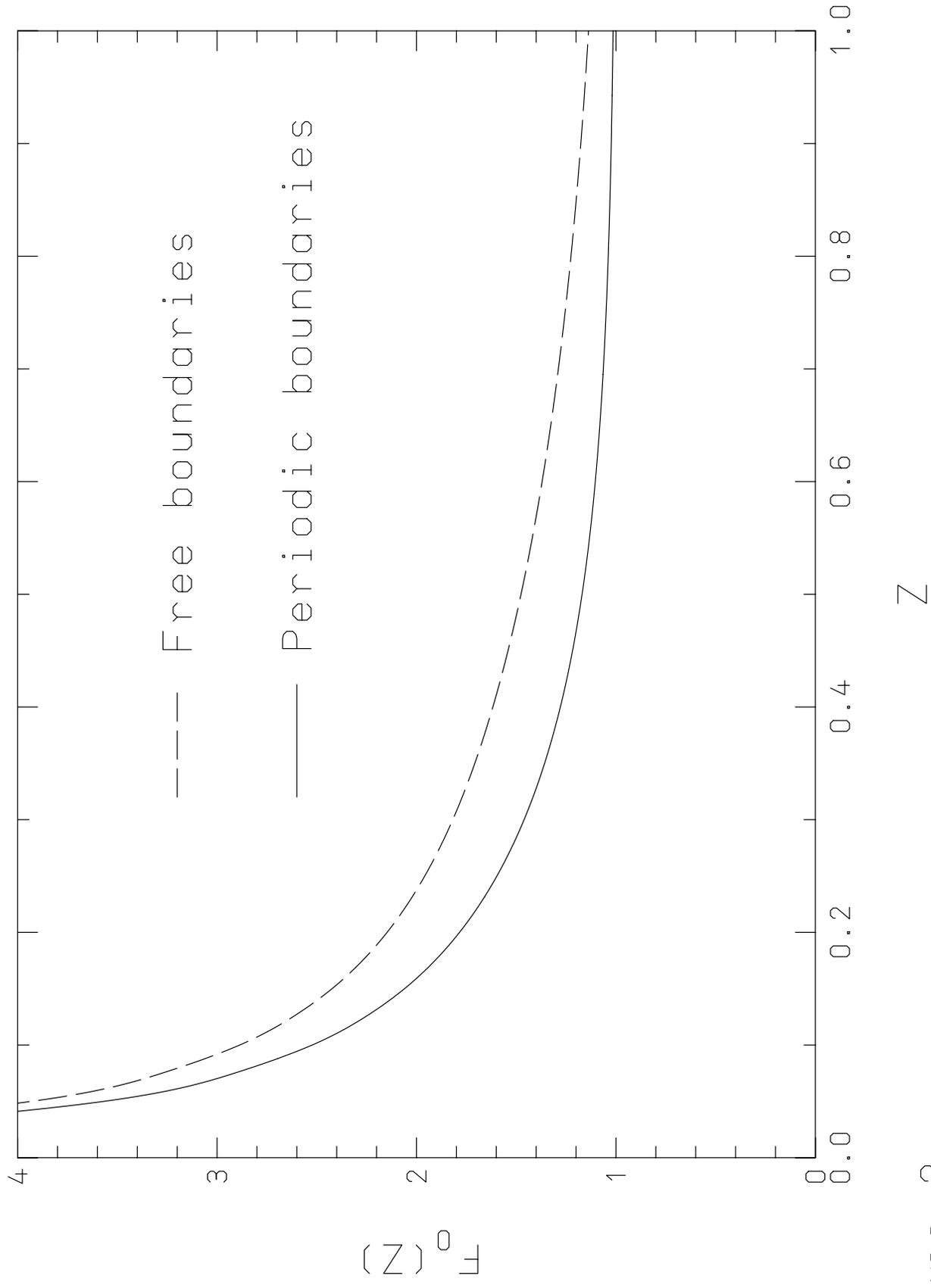
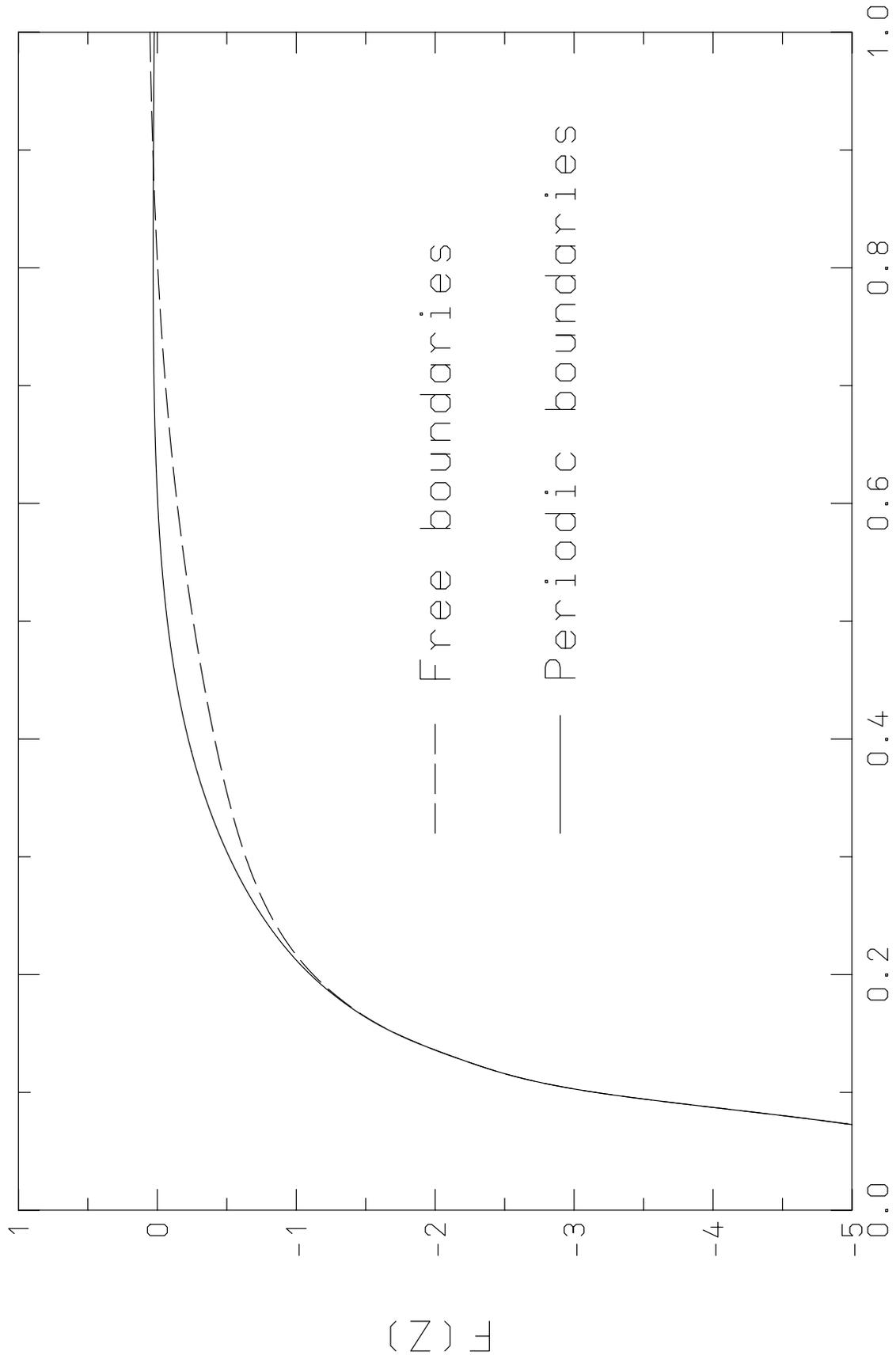
\end{document}